# Spectrum of Solar Type I Continuum Noise Storm in the 50 - 80 MHz band, and Plasma characteristics in the associated source region

G. A. Shanmugha Sundaram[1,2]

*Joint Astronomy Programme, Department of Physics, Indian Institute of Science, Bangalore - 560012, Karnataka, INDIA*

sga@physics.iisc.ernet.in

shs@iiap.ernet.in

and

K. R. Subramanian[2]

*Indian Institute of Astrophysics, Koramangala, Bangalore - 560034, Karnataka, INDIA*

subra@iiap.ernet.in

## ABSTRACT

Continuum observations of a solar noise storm in the frequency range of 50 - 80 MHz observed with the Gauribidanur radio spectrograph during 2000 September, 26 & 27, are presented here. The radio spectral index of the noise storm continuum in the band 50 - 80 MHz is found to be $\approx 3.65$ during the above period. The Noise Storm continuum radiation is explained as a consequence of the non-thermal, plasma emission mechanism. The beam-density of suprathermal electrons is estimated for the coronal plasma near the source region of storm radiation. Supplementary evidence for the density-estimate is provided by way of analysing the imaging data from the SXT on-board the Yohkoh spacecraft, and the LASCO, MDI, and EIT on board the SoHO spacecraft.

*Subject headings:* Sun: radio radiation — radiation mechanisms: non-thermal — plasmas

---

[1]Joint Astronomy Programme, Department of Physics, Indian Institute of Science, Bangalore - 560012, Karnataka, INDIA

[2]Indian Institute of Astrophysics, Koramangala, Bangalore - 560034, Karnataka, INDIA

## 1. Introduction

Solar noise storms are one of the most studied phenomena, since their discovery in the year 1946 (Hey 1946). They consist of a broadband ( df/f $\approx$ 100% ) continuum lasting for hours to days, superimposed with narrow-band ( df/f $\approx$ 3 % ) short duration ( 0.1 - 1 s ) radio bursts. The continuum components of the Type I noise storms mostly occur from 30 to 300 MHz, and their intensity exhibits a slow variablity with time. The study of the flux density, brightness temperature, and size of the noise storm continuum are important from the point of view understanding their emission mechanism. The brightness temperature of the continuum was found to increase with frequency, with a brightness temperature value ranging from $5 . 10^8$ K at 43 MHz (Suzuki 1961) to about $9.4 . 10^9$ K at 169 MHz (Kerdraon 1973).

Multifrequency observations of the noise storm continuum events (Thejappa and Kundu 1991, a), below 80 MHz, have determined the brightness temperature spectral index to be positive. We present the flux density measurements of a noise storm continuum in the 50 - 80 MHz band observed on 2000 September, 26 & 27, and derive the spectral index. Plasma emission mechanism is suggested as the explanation for the estimated spectral index value. With a knowledge of the continuum source size determined from the spatially resolved radio imaging data of the Gauribidanur and Nançay RadioHeliographs, the brightness temperature, and hence the beam-density of the suprathermal electrons in the source region are estimated. Additional evidence for spatial counterparts of the noise storm source region, and their associated plasma parameters at the photospheric and coronal levels, is provided from analysis of the imaging data from the Yohkoh Soft X-Ray Telescope ( Y-SXT ), $H_\alpha$ image from the Big Bear Solar Observatory, and the Large Angle and Spectrometric Coronagraph ( LASCO ), the Michaelson-Doppler Imager ( MDI ), and the Extreme ultraviolet Imaging Telescope ( EIT ) on board the SoHO spacecraft.

## 2. Observations

The observations presented here were made with the Gauribidanur radio spectrograph ( GRS (Ebenezer 2001)). This spectrograph operates in the frequency range of 30 - 80 MHz with a frequency resolution of 1 MHz and time resolution of 3.3 seconds. This spectrograph is used in conjunction with 8 log periodic dipoles of one group of the Gauribidanur radio heliograph (Ramesh et al. 1998). These dipoles can operate in the frequency range of 30 - 150 MHz. The system has the sensitivity to record the quiet sun, and strong radio sources Cyg.A and Cas.A as point sources. The spectrograph is used to observe the sun during the meridian transit. There was an intense noise storm during the period 2000 September, 20

- 30. Observations of the noise storm continuum in the band 30 - 80 MHz were made on 2000 September, 26 and 27. Figure 1 and Figure 2 show the drift scans of the noise storm continuum emission at selected frequenies. The data samples at each frequency are separated by 3.3 seconds.

## 3. Data Analysis

In order to determine the peak amplitude deflection of the solar scans at different frequencies, 60 data-points ( 3 minutes duration ) around the time of transit were chosen. For the baseline determination 50 data points( 2.5 minutes duration) were chosen and median-filtered to remove any isolated high or low values in the baselines variations. The difference between these two deflections gives the uncalibrated amplitude.

Absolute flux calibration of the scans at different frequencies was performed using the radio source Cygnus A ($19^h59^m33.^s0$, $+40°43'41.''0$). The flux density of Cygnus A, at frequencies in the band 30 - 80 MHz, was derived using the the expression for the radio flux of Cygnus A ((Baars et al. 1977)):

$$\log S = [\ 4.695(\pm0.018)\ ] + [\ 0.085(\pm0.003)\ .\log\nu\ ] + [\ -0.178(\pm0.001)\ .\log^2\nu\ ] \quad (1)$$

where, S is the radio flux of Cygnus A in Jansky, $\nu$ is the frequency in MHz, and the above equation holds good in the frequency range of 20 MHz to 2 GHz. Data at fourteen frequency channels were found to be suitable for the present analysis. A gain correction factor of 1.8 dB was applied to deflections of the Sun on both days, to offset the variation in gain of the array-pattern with differing declinations of the Sun and the calibrator.

The contribution of the Quiet Sun component of the radio flux was subtracted from the calculated fluxes at 14 frequencies. The Quiet Sun flux was determined using the expression given by (Erickson et al. 1977).

$$\log S_\odot = -0.2\ (\pm0.21) + 2.25\ (\pm0.12).\log\nu \quad (2)$$

where, $S_\odot$ is in Jansky, and the frequency $\nu$ is in MHz.

The solar scans at different frequencies now consist of the contribution due to both the bursts and continuum. The contribution due to radio bursts was removed by the following techniques used in (Malik and Mercier 1996). The differentiation of the data with respect to time was done for 60 points about the transit. A burst shows a steeper slope compared to the continuum. Data from time periods that have no steep change in the slope (i.e. $dS/dt \geqslant 0.1$ ) were used for further analysis of the continuum. The next step was to evaluate the distribution of intensities around the transit point for the 3 minutes duration. If 50 % of the points lie below the peak value of the distribution, then those data points were used in the estimation of the continuum flux. The errors in the estimated flux density of the continuum are $\lesssim 10\%$.

In Figure 3, a log-log scatter plot of the frequency vs. the noise storm continuum flux is shown, with a Chi-squared polyfit. The slope of the line gives the value of the spectral-index, and is 3.71 on 26th & 3.65 on 27th of September, 2000.

In order to determine the brightness temperature, the continuum source size was estimated from measurements at several other frequencies. Table 1 is a compilation from published data ((Bougeret 1973), (Kerdraon 1973), (Steinberg Caroubalos and Bougeret 1974), (Bougeret and Steinberg 1977), (Elgarøy 1977), (Melrose 1980, a), (Melrose 1980, b), and references therein ) on continuum source sizes, and includes the source sizes determined from imaging observations of the Sun with the Gauribidanur Radio Heliograph (GRH) at 109 MHz and the Nancay Radio Heliograph (NRH) at 164 MHz, on 2000 September, 26 and 27.

A plot of the continuum source size with frequency is shown in the Figure 4, along with the fitted falling-exponential curve of the form,

$$D = 293.5 . (1 - exp(-\nu^{-0.75})) \quad arcmin. \qquad (3)$$

where, $\nu$ is the observing frequency in MHz and D is the noise-storm half-power source-size in arc min. The falling-exponential curve shows a high degree of correlation ( Corr. coeff.: 0.989 ) with the data points listed in Table 1, especially those in the frequencies upto 190 MHz, which is well within the upper limit to the frequency-range of the GRH ( viz., 150 MHz ). The curve also fits to a high degree, the source-sizes, of the noise-storm event considered, obtained from the imaging data of the GRH ( 109 MHz ) and the NRH ( two data-points at 164 MHz ), as would be explained in Sec.(5.1). The value of half-power source size is $\sim 8.'7$ at 109 MHz, and was derived from the Gauribidanur Heliographic observations. The two data-points at 164 MHz correspond to continuum source sizes measurements obtained from the two-dimensional RadioHeliographic maps of Nancay, posted in their website. The source-size of the noise-storm events, at the 14 frequency channels between 50 MHz and 70 MHz, encompass a smoothing-factor caused by the large beam-width of the GRS, and are well represented by Eq. (3), for the spectrographic observations done on the same two days. The value of half power source size is $\approx$ (13.2 $\pm$ 1.2 ) arc min. in the 30-80 MHz band.

The brightness temperature is given by the expression:

$$T_b \; = \; \frac{S \cdot \lambda^2}{2k\Omega} \qquad (4)$$

where, S is the continuum radio flux in Jansky, k is the Boltzmann's constant, $T_b$ is the brightness temperature in $deg$. K, $\Omega$ is the continuum source-size in arc minutes and $\lambda$ is the wavelength in meters. The derived $T_b$, varies from $1.07 \cdot 10^7 \; ^{\circ}K$ to $1.96 \cdot 10^7 \; ^{\circ}K$ in the 50 - 80 MHz frequency band. From a knowledge of the $T_b$ of continuum noise-storms at a particular frequency, the number density of the suprathermal electrons, that are trapped in the associated magnetic flux-lines, can be estimated. The values of spectral index ($\approx +3.7$) and $T_b$ ($\approx 10^7$ K ) strongly suggest a non-thermal plasma emission mechanism.

## 4. Discussion

According to the current models of noise storms, the radiation is believed to be due to plasma emission. The coalescence of Langumir waves excited by the suprathermal electrons with low frequency ( ion - acoustic or lower-hybrid waves) converts the Langumir waves to electromagnetic radiation. (Melrose and Stenhouse 1977), (Thejappa 1990), (Thejappa and Kundu 1991, a), and (Thejappa 1991, b) showed that the observed brightness temperature

$$T_b \; = \; (\alpha^L \; / \; \gamma^L) \qquad (5)$$

is related to the effective temperature of the Langumir waves. Here, $\alpha^L$ and $\gamma^L$ are the emission and absorption coefficients respectively (Melrose and Stenhouse 1977). The absorption coefficient is the sum of the collisional frequency $\nu_c$, the negative damping due to trapped particles $\gamma_A$ and the Landau damping by ambient coronal electrons $\gamma_L$. Hence,

$$T_b \; = \; \alpha^L \; / \; (\gamma_A \; + \; \nu_c \; + \; \gamma_L) \qquad (6)$$

(i.e.),

$$T_b = \frac{0.3 \cdot (n_b/n_e) \cdot \omega_p \cdot m_e \cdot V_{T_b}^2}{-0.04 \cdot (n_b/n_e) \cdot \omega_p + \nu_c + \gamma_L} \qquad (7)$$

In the above expression for $T_b$, $n_b$ is the electron beam density, $\omega_p$ is the plasma frequency, $V_{T_b}$ (assumed to be equal to the phase velocity $v_{ph} = 10^{10} \ cm \ s^{-1}$) is the thermal velocity dispersion of the trapped particles, $m_e$ is the electron mass, and $(n_b/n_e)$ is the density ratio of suprathermal electrons, the threshold value of which $(n_{th})$, defined for the limiting case where collisions are balanced by negative-damping, leaves the effective damping to be solely contributed by Landau damping phenomenon of the background electrons.

The primary factor responsible for the emission of Type I noise storms ( T1NS ) is a trapped electron- distribution, and the effective temperature $(T_{eff})$ of the L-waves is equated to the observed $T_b$, by invoking the case of an isotropic electron-distribution and spontaneous L-wave emission, to explain the noise storm continuum. According to (Thejappa 1991), the limiting value of $T_{eff}$ and hence $T_b$, for the steady L-wave emission required in the case of noise storm continuum, is $10^{11}$ K, below which $\nu_c > -\gamma_A$. If the source region is assumed to be stationary and homogeneous, $T_{eff}$ is equal to the effective temperature of L-waves ( $T^L$ ) only when the absorption coefficient is much greater than the source size. $T_{eff}$ tends to a maximum value of $T^L$, in the limiting case for an optically thick source region, with respect to the decay of the transverse waves to the L-waves and the lower-hybrid waves, thereby signifying the threshold. Hence, determining $T_{eff}$ involves the estimation of $\alpha^L$ & $\gamma_L$. As the density of the suprathermal electron beam increases such that $\nu_c < -\gamma_A$, the emission flips from an incoherent to a coherent one, and this explains the high brightness temperatures associated with intense Type I bursts.

The threshold density of the suprathermal electrons is a decreasing function of frequency, for a particular storm at a given instant; a slight fall in the value of ( $n_b/n_e$ ) violates the conditions for threshold, causing $T_b$ to become equal to the temperature of the background continuum.

For the case of Eq.(7), when the collisional frequency and Landau damping coefficient have values of $\nu_c = 1.25 \ s^{-1}$ & $2.01 \ s^{-1}$, and $\gamma_L = 3.10^{-2} \ s^{-1}$ & $3.7.10^{-2} \ s^{-1}$, the threshold values for the density ratio $(n_b/n_e)$, at the extreme frequencies considered (viz.), 54 MHz and 69 MHz, are $1.10^{-7}$ and $1.2.10^{-7}$ respectively. These are in general agreement with the values published earlier (Thejappa 1990), (Thejappa and Kundu 1991, a), and (Thejappa 1991, b).

Assuming a density enhancement factor of 2.5 over that of the Newkirk coronal electron density model (Newkirk 1959) for the active region of the corona associated with the noise storm, the values of electron beam density of suprathermal electrons ($n_b$) range from 3.4 to 6.95 $cm^{-3}$ in the 54 to 69 MHz frequency range.

The estimation of $T_b$ is made feasible by knowing the threshold beam-density for the case where $\nu_c$ balances $-\gamma_A$. Hence, Eq.(7), which incidentally constitutes the solution for the radiative-transfer equation, holds good only for a optically thick source region, where, ($l > \Delta r$) and ($\nu_c + \gamma_A = 0$), & ($\gamma^L = \gamma_L$).

For values of $v_{ph} \approx v_{T_b} \approx 10^{10} \ cm \ s^{-1}$ and $v_T \approx 3.89 \cdot 10^8 \ cm \ s^{-1}$, the absorption length ($\Delta r$) has values ranging from $2.13 \cdot 10^{-2} \ R_\odot$ to $1.75 \cdot 10^{-2} \ R_\odot$, while the source-size (L) of the noise-storm event chosen for the work, according to Fig.(4) and Eq. (3), varies from 0.9 $R_\odot$ to 0.75 $R_\odot$ for the 14 frequency channels chosen in the 50-70 MHz band. Hence, the condition for an optically thick case ( $L > \Delta r$ ) is easily satisfied, and Eq. (7) gives a valid estimate of brightness temperature of the continuum noise-storm event.

The inhomogeneous nature of the corona leads to refraction, diffusion, and scattering involving the L-waves and the noise-storm radiation. The values of $T_b$ estimated here are bound to fall short of the actual value of continuum brightness temperature values. A quick verification of the $T_b$ values obtained from the dynamic-spectral studies, made feasible by examining the Soft X-ray and extreme ultra-violet imaging observations, would reveal the extent of offset from the actual value of $T_b$, and a measure of the scattering processes involved. A density analysis of the noise storm source regions that occur above the SXR and EUV coronal loop-structures, has found enhancements that are a factor of $\approx 10$ more than the ambient coronal densities estimated for fundamental plasma emission (Stewart Brueckner and Dere 1986). A density-factor of 2-3, assumed for the study of plasma parameters from the spectral data on noise storms, can be verified from a similar analysis on the Soft X-ray and extreme ultra-violet data. The next section describes such an analysis done on the Yohkoh-SXT and SoHO-EIT images, and the radio-imaging data from GRH and NRH, in addition to the supplementary evidence, relating to magnetic-flux changes in the background active-regions, obtained from the SoHO-MDI. The procedure would also validate the assumption of the density-enhancement factor chosen over that of the Newkirk coronal density model, and the presence of associated coronal streamers at the sites of noise-storm activity ( as verified from the SoHO-LASCO C2 Coronographs ).

## 5. Supplementary evidence

Type I noise storms are known to be intimately linked to photospheric bipolar sunspot activity ( (Hey 1946; Bentley et al. 2000)). The reconnection of newly-emerging magnetic flux lines with the pre-existing ones leading to the release of radio flux by way of expending the magnetic free energy (Benz and Wentzel 1981) is being cited as a possible reason for noise storm activity. This is observed as a likelihood for noise storm occurance as a delay to the photospheric changes in the active regions associated with storm activity. There exists a greater possibility for the suprathermal electrons, which cause the T1NS, to remain trapped in these magnetic field structures above the active regions, and in turn exciting the upper hybrid waves in the ambient plasma. The interaction of such high frequency waves with the ion-acoustic waves, occuring at the site of the current sheet in the reconnection region is being proposed as the alternative possibility for noise storm radiation. In such cases, the continuum source would be located on the same magnetic field lines. Another probable cause for noise storm emission may be the role played by the super- Alfvénic shock waves propogating through a region of newly-emerging magnetic fields, and the resultant energy (Spicer Benz and Huba 1981; Benz and Wentzel 1981). The continuum emission is thus rooted in the supra-thermal electrons at the shock wavefront, which cause a localized transfer of momentum in the plasma, leading to microturbulence (McLean 1981) and noise storm excitation. The energetic non-thermal population of electrons, which cause the metric noise storms, have a coherent emission mechanism. It would be of considerable interest to look for its metric radio wavelength signatures along with associated activity in solar magnetograms, Long-Duration Soft X-ray events, and UV data , in line with the previous works ( (Krucker et al. 1995; Bentley et al. 2000; Lantos 1981; Kerdraon et al. 1983)), for this particular period, to corroborate / determine the results on electron density, plasma emission-measure, brightness temperature, source-size, and altitude of the source above the photosphere, relative positions of the optical, x-ray, and extreme ultra-violet counterparts to that of the noise storm source region.

Estimation of the various solar coronal plasma parameters was made, after careful analysis and reduction of data obtained from the BBSO, the Yohkoh-SXT, and the LASCO, MDI, & EIT payloads on board the SoHO spacecraft, using the integrated software libraries, databases, and system utilities, functioning primarily under an IDL (URL1) based Solaris operating environment termed *SolarSoftWare* ( SSW (URL2; Bentley and Freeland 1998; Freeland and Handy 1998)).

## 5.1. Radio Imaging with the GRH & NRH

The radio imaging data of Sun, on 26, 27 Sep. 2000, as depicted in Table 2 and Fig. 5, have been obtained from radio observations made using the Gauribidanur and Nançay RadioHeliographs.

### 5.1.1. Gauribidanur RadioHeliograph

The Gauribidanur RadioHeliograph (GRH), in its current configuration, images the Sun at a frequency of 109 MHz. The HPBW of the heliograph array is ( $5.'4 \times 7.'8$ ) along the (E-W) × (N-S) (Ramesh et al. 1998). The CLEAN algorithm is used to obtain radio maps of the Sun, as it transits the local meridian of Gauribidanur. A snapshot image of Sun, done in an integration time of 256 ms at transit, has a dimension of $103 \times 103$, and the plate-scale is $1.'165$. The total field-of-view for the radio-image is ($120' \times 120'$). The region of noise storm source, as identified from the GRH image, has a half-power size that is the convolution of the power pattern of the GRH array and the actual size of the source. Since the (E-W) beam size is narrower, the source region of the noise storm, as resolved by the GRH beam, has a post-deconvolved size of $8.'7$, along the Solar (E-W). Absolute radio flux calibration was performed by observing the intense, unresolvable ( at the operating frequency of the GRH ) radio source Cygnus A.

The calibrated solar image on 27th September, 2000, whose flux contour-plot is depicted in Figure.(5), has a peak flux of 89 sfu, and the adjacent equispaced contour levels are apart by 9.74 sfu. The corresponding cross-section of the main-beam of the heliograph array at 109 MHz is shown on the lower-left corner of the image.

### 5.1.2. Nançay Radioheliograph

The Nançay Radioheliograph (NRH), located at 47 N : 02 E ( NRH (RH Grp.) ) consists of a T-shaped array of antennas mapping the Sun at five frequencies between 450 and 150 MHz (0.7 m - 2 m wavelength) with sub-second time resolution (Kerdraon and Delouis 1997). Radio emission at these frequencies originate from the low and middle corona (height range roughly 0.1 - 0.5 solar radii above the photosphere). The beamwidth at 164 MHz is $1.'2$ along the E-W and $\approx 3.'1$ along the N-S.

The imaging data used for this particular work, consists of upto 5 images obtained for each of the days 26 and 27 Sep. 2000. Radio images, obtained during transit of the Sun across the local meidian, have been used because of relatively less distortion caused to the incoming radio signals by the ionosphere at the instrument zenith. The images are available in the public domain of NRH in the FITS data format, and constitute a $(256 \times 256)$ matrix, with each pixel measuring 1.56 % of $R_\odot$ along one dimension. The high angular resolution enables accurate estimation of the coordinates of the noise storm centroid, in relation to the imaging data obtained from other wavelengths. Figure.(6) is a $(2R_\odot \times 2R_\odot)$ radio-map obtained with the NRH at 164 MHz on 27th September 2000, 10:56:14.82 UT. The small oval on the lower-left corner of the image is the beam-size of the NRH array at 164 MHz, and the radio-source has been colour-contoured. The peak-flux is estimated to be 432.8 sfu, and adjacent coutours are separated by (6l) sfu where, l=1,2,3,... Table 2 is obtained from the monthly catalogue of noise storms at 164 and 327 MHz (SGD-092K). The relative positions of the noise storm centroid as a function of frequency are evident from the data for the two days considered.

The estimated peak-flux values for the noise-storm continuum event on the 27th of September 2000, at frequencies of 109 MHz & 164 MHz, by the GRH & NRH respectively, are 89 & 432.8 sfu. This information, obtained from the 2-D imaging data, can be applied to determine the continuum spectral-index, using the expression :

$$(S_G/S_N) \;=\; (\nu_G/\nu_N)^\alpha \tag{8}$$

where, $S_G$ & $S_N$ are the peak-flux values at the operating frequencies $\nu_G$ & $\nu_N$ of the GRH & NRH respectively, and $\alpha$ is the spectral-index. The value is found to be $\approx +3.8$, and is in remarkable agreement with the $\alpha$ determined from dynamic spectral records of the continuum noise-storm event.

### 5.2. BBSO H-Alpha image of AR complex

Emergence of magnetic flux in bipolar magnetic region ( BMR ), and their subsequent inter-connections with distant BMRs (Brueckner Patterson and Scherrer 1976) constitute an ideal environment for noise storm radiation. The photospheric features associated with the Type I Noise Storms on 2000 September 26 and 27, are the active regions AR9169 and AR9170. The centroid of the latter made the Central Meridian Passage ( CMP ) on Sep.24, at the heliospheric latitude of $5^o$ $S$. The former had the CMP on Sep.24, and was located at $9^o$ $N$. Table 3 lists out the details on each of the active regions, sourced from the SGD (SGD-092K). The leading spot in the northern hemisphere had positive polarity, while the trend was the opposite for the AR on the south. The area mentioned in the table is in units that are a millionth of the area of the solar disk. The values of area and number of sunspots in the ARs are well above the necessary minimum requirement of 100 millionth of the solar disk size (Dodson and Hedeman 1957; Dulk and Nelson 1973), thereby amply associating them with the position and period of the noise storm event in perspective.

The Full-disk H-alpha image, as observed by the Big Bear Solar Observatory ( BBSO ), on 2000 September 27, at 14:57:47 UT, with an exposure time of 30 ms, is depicted in Figure.(7). The image has been dark-subtracted, flat-field corrected, and limb-darkening subtracted.

The ARs in context evidently have a complicated, multipolar configuration, suggestive of localized neutral current sheet in the corona , and a magnetic neutral line in the underlying photosphere : such regions tend to be strongly linked to intense noise storm activity, as would be described vividly in the following sub-section.

### 5.3. Coronal streamers detected by the LASCO Coronograph

The beam of suprathermal electrons that contribute to the continuum noise-storm emission at the fundamental plasma frequency, propagate radially outward from the Sun, along the coronal open magnetic field-lines. Since the direction of the radiation wave-vector coincides with the direction of the streaming electron beam, the L-waves travel radially outward as well. The location of the observed continuum source is also the site for a complex magnetic geometry arising out of the reconnection of the field lines among active regions. The optically identifiable feature of such an electron stream threaded by open magnetic fields is the coronal 'helmet'-streamer, also referred to as the *active-streamer* ( (Elgarøy and Eckhoff 1966) & references therein ), positioned above sites of localized, intense photospheric magnetic fields, and extending out, as density-enhanced features in coronograph images, to distances of 1.5 - 2 $R_\odot$. The continuum source would be situated on the vertical current sheet, about which the reconnection phenomenon occurs. The electron density of the streamer would typically vary from 2-5 over that of the Newkirk coronal density model.

The Large Angle and Spectrometric Coronagraph ( LASCO ) instrument is one of 11 instruments included on-board the SoHO (Solar and Heliospheric Observatory) spacecraft. LASCO is a set of three coronagraphs that image the solar corona from 1.1 to 32 solar radii, of which the C2 coronagraph has a central occulting disk extending from 1.5 to 6.0 $R_\odot$ ((SoHO Sp.; URL3; URL4), and references therein).

The LASCO C2 Coronogram was used to identify the presence of a coronal streamer at the position-angle where the noise-storm source had been located, and hence infer the electron density in the region occupied by the streamer. The C2 Coronograph FITS data was made available in compressed telemetry format, termed the *Level-1* data-set. This had to be decompressed, bias, and stray-light ( due to cosmic rays ) subtracted, dark-current & flat-field corrections performed, and the geometric-distortion, vignetting, warp, and missing-blocks corrected ( where the information on lost blocks at the high spatial-frequencies is retrieved using the fuzzy logic method ) suitably; the entire procedure constitutes the standard pipeline-processing, to take the Level-1 image to *Level-2*, suitable for quantitative scientific analysis. This step is followed by calibration in units of Mean Solar Brightness ( MSB ). Unit MSB is $2.01 \times 10^{10}$ *erg* $s^{-1}$ $cm^{-2}$ $sr^{-1}$, and is defined for the mean Sun-Earth reference distance of 1 AU. The solid angle subtended by the Sun's disk at the location of the SoHO spacecraft, positioned at Lagrangean L1, is $\approx 8.9 \times 10^{-5}$ sr. The electron density per pixel is determined from the Thomson electron scattering equations. The total brightness due to the target electron, whose impact-distance and distance from sun center (in $R_{\odot}$), and Polarization properties are known, is thereby computed. A limb-darkening factor of 0.63 is included to account for effects due to curvature on brightness per pixel.

The Coronogram has dimensions of ($1024\,pix. \times 1024\,pix.$), and the central occulting disk renders a view that extends radially outwards from $(1.5-6)R_{\odot}$. The pixel-size is $\approx 11.\!''4$. Fig. 8 is the LASCO C2 Coronograph image obtained on the 2000 September 26, at 08:06:05.884 UT. The image shows the occurance of a coronal streamer on the N-W limb. The Preliminary 2000 SOHO LASCO Coronal Mass Ejection List (LASCO-CME List), compiled based on probable CMEs detected in white light observations of the LASCO coronagraphs' quick-look data (*Level-0*), mentions about the event as a large, bright ragged loop with core occuring on the NW. The description fits the morphology of the helmet streamer, and the position correlates with that of the noise storm source region as in Table 2.

The visible extent of the streamer in the C2 coronogram is $1.5-2.82R_{\odot}$, and the estimated electron density is about $5.13 \times 10^7$ at distances of $1.55-1.84R_{\odot}$; this corresponds to a density enhancement factor of (2 - 1.23), for the fundamental plasma emission in the 54 - 69 MHz range. The value for the density enhancement factor, is nearer to the assumption of d=2.5 made in the previous section, in order to determine the threshold density for suprathermal electrons in the continuum noise storm source region. The corresponding intense feature on the N-W limb, is a result of the active streamer forming an arc- like structure in the lower corona, above the large, complex sunspot group constituting ARs 9169 and 9170.

## 5.4. Full-Disk Magnetogram / MDI

The Michelson-Doppler Imager ( MDI (Scherrer et al. 1995) ) measures the Zeeman splitting of the Ni I line at 6767.8 $A^o$. Longitudinal magnetograms represent the Doppler shift of this line, observed with two opposite circularly-polarized wave plates; a difference in the two constitutes a measure of the Zeeman splitting, and in turn the magnetic flux-density of the mean line-of-sight component of the magnetic field over the imaging picture-element. In Fig.(9, 10 & 11), the MDI F-D synoptic magnetograms, each of image-dimension (1024 $pix.$ × 1024 $pix.$), obtained with a cadence of 96 Min. and a spatial resolution of 1."98 are shown, overlaid with the contours of the solar noise storm images obtained with the NRH and the GRH. The accompanying color scale-bar indicates the relevant polarity on the magnetogram. Table(4) gives details on the active regions, that are also observed on the magnetograms, on days 2000 September 26 and 27.

Noise storms have been known for their association with active-regions and bipolar sunspot groups on the solar disk; their onset apparently leads to the formation of extended, dense loops in the mid-coronal altitudes, connecting ARs that are great distances apart, while alluding to the eventuality of large-scale restructuring of the coronal and photospheric magnetic fields (Stewart Brueckner and Dere 1986; Bentley et al. 2000). There exists a unanimity in the view that magnetic fields occur in the photospheric features associated with the active coronal regions, yet opinions differ on the estimated magnetic field strength that prevails there. Estimates on magnetic field strength, as observed from the magnetograph data in the $H_\beta$-line (Korol'kov and Soboleva 1962), arrive at values of 300-500 Gauss above the sunspots, while field strengths of 1000-3000 Gauss are found at photospheric levels, below the coronal condensates. Hence an analysis of the magnetograms was attempted, in order to estimate the magnetic field strengths, and to determine whether any fresh emergence of magnetic flux, as evidenced from an abrupt change in the magnitude or sign ( flux-reversal ) in the magnetogram data, had occured. Noise-storms are known to occur, at sites above the flux-tubes emerging out of the photosphere, high-up in the corona, in the 30 - 400 MHz plasma-frequency range, due to reconnections of magnetic flux lines, either with pre - existing lines of magnetic flux, or with newly-emerging magnetic flux ( that perturb the overlying magnetic fabric and render them unstable ), with their foot-points located at the sunspot-groups of unlike polarity. The reconnection process tend to migrate to the higher reaches of the corona, while the footpoints spread farther apart; a rough estimate is that the loop-height would be one-half of this separation. The spreading-apart is a result of the footpoints getting constantly repaired to sites in the AR complexes that have new BMRs and flux emergence. The loops grow larger due to several reconnections with the older ones, constantly forming new loop systems. The polarity of the noise storms, that are located above the loop-regions, is the same as that of the dominant polarity of the spot-region that's in the backdrop.

The plot of mean magnetic flux per pixel ( in Mx/DN ) ( the mean magnetic flux-density or MMFD ) of the MDI-FD magnetogram image, from the 23rd to the 28th of September 2000, is shown in Fig. 12. With 15 images each day, a total of 90 magnetograms ( or 144 hours ) were analyzed for the six days chosen. The method of analysis of the image-data involves choosing a particular active-region ( whose coordinates in the solar-disc were obtained from the Solar Geophysical Data Reports ), performing an average along the solar latitude and solar longitude for the rectangular-window chosen, and study the variation of the flux - density value ( whose units are in terms of Mx / DN ) for the period. The window, whose dimensions are held constant for the entire period considered, was moved across the disc at such a rate that, the active-region occurs at exactly the median- position; the rate is corrected for differential rotation and solar- curvature. Since this rectangular window also includes portions of the quiet solar-disc, in order to excise their influence on the mean flux value, a separate region was chosen on the disc, farther away from any active-region ( and preferably closer to the poles ) on a case-by-case basis for each image, and the relevant mean-flux per pixel was determined; this in turn was subtracted from the value for the active- region obtained as an average, so that the resultant value has a contribution to the magnetic-flux density from the active-region alone.

The catalogue of Outstanding Solar Radio Emission Occurances (SGD-0301) lists the dynamic spectral data of San Vito ( SVTO ) with the noise storm characteristics as shown in Table 4. With reference to Fig. 12, the MMFD reached values of $6.76 \times 10^{13}$ and $6.97 \times 10^{13}$ for the magnetograms of 9:36 UT and 14:24 UT. on the 26th, as against the low values for MMFD observed during the periods preceding the noise-storm event, suggesting a strong temporal correlation of the peak noise-storm radio-flux with the MMFD.

The MMFD values were about a third on the 27th, in comparison to their values the previous day; yet, the first three entries for in Table 4 suggest long duration noise storm activity right from the early part of the day. This is corroborated by the MMFD values depicted in Fig. 12 by the first peak on the 27th. The final entry in Table 4 shows the noise storm flux as detected by Palehua ( PALE ) at 245 MHz. The peak radio flux of 310.0 sfu at 18:01 UT for the noise storm, corresponds to the three successive magnetograms, with the peak MMFD observed at 17:36 UT.- the second peak on the 27th.

From the foregoing discussions, it's amply convincing that a hike in MMFD in the associated active region is temporally succeeded by a similar peak-flux in the noise storm emission, while a notable future at the peak noise storm flux epoch is one of polarity reversal of the magnetic-flux, for the pixel being tracked. This, along with the spatial association of the noise-storms, as shown in Fig.(9, 10 & 11), suggests that the rearrangement of magnetic flux at the sites of active regions has a definite role to play in initiating and sustaining the associated noise storm activity in the corona.

## 5.5. Coronal Loops / EIT

Noise storm sources are known to occur above extended coronal loop-structures as observed in the extreme ultravoilet, with the footpoints of the loop traced down to magnetic regions on the photosphere (Sheeley et al. 1975, a), (Sheeley et al. 1975, b). In order to investigate such a phenomenon, imaging data from the EIT was studied. The Extreme Ultravoilet Imaging Telescope ( EIT (Delaboudinière et al. 1996) ) is a normal-incidence, multilayer telescope carried on-board the SoHO spacecraft. Full-disk imaging of the Sun is done at four, selected bandpasses in the extreme uv, and their image-characteristcs are shown on Table 5. The field of view is $45' \times 45'$ ( or about 1.3 $R_\odot$ at Lagrangian L1).

The EIT data, on days 26th and 27th of September 2000, was analyzed in all four filter bands, to estimate the temperatures and electron densities (Emission Measures) in regions that have a high degree of spatial correlation to noise storm activity. The following procedures were performed to convert the data-type ( obtained in compressed telemetry format - termed Level-1 EIT data ) associated with each pixel in terms of physical quantities.

- The values associated with missing pixel blocks was fixed either to those that were of a darker corner in the same image or with the average value of the surrounding blocks.

- Degridding to excise-out the contribution made by the filter grid pattern in the image.

- Background-noise ( dark-current ) subtraction, flat-fielding, vignetting, and degradation corrections.

- Computing the normalized filter response for each of the four EIT bands; the resulting normalization factor is applied to the measured DN level on any given date.

- Image calibration using valid averaged calibration lamps obtained as an updated look-up table.

The *CHIANTI* ( an atomic database for spectroscopic diagnostics of astrophysical plasmas (URL0) ) database on emission line spectral data was used to compute the synthetic spectra ( units = $photons\ s^{-1}sr^{-1}\ cm^3$ or ergs.) (Mewe Lemen and van den Oord 1986), and hence the Column Emission Measure, electron density, temperature, and element abundances, as an isothermal emission measure approximation. Assuming a Solar Emission Measure of $10^{44}\ cm^{-5}$ at L1, the brightness temperature and EM were estimated for a given ratio of either (FeXII/FeIX,X) or (FeXV/FeXII). In Fig. 13, the EIT image is overlaid with the GRH contours. The centroid of the noise storm source region lies near to the site of large-scale coronal magnetic arches ( trans-equatorial loops ) connecting the ARs on either hemisphere. The EIT image was obtained with the 195.127 $\mathring{A}$ (Fe XII) filter, on 2000 Sep. 27, at 23:36:11 UT. The exposure time was 2.5 s. The image is of dimensions $(512 \times 512)$, with a plate-scale of 5.″24 . The coronal loop structure on the N-W limb, at the vicinity of the noise-storm source, had a thickness of $d = 1.89 \times 10^9$ cm. The EM varies from $(0.98 - 1.3)\ \times\ 10^{29}\ cm^{-5}$, and the brightness temperature is $1.68\ \times\ 10^6$ K. The emission measure per unit area of the loop normal to the magnetic field lines is given by the expression :

$$EM\ =\ \int_{source} n_e(l)^2\ dl\ \approx\ <n_e>^2 qd\qquad cm^{-5} \qquad (9)$$

In Eq. 9, with a loop filling-factor ( q ) assumed to be $\approx\ 1$, and applying the EM value determined from the EIT data, the mean electron density in the region closer to the noise storm works-out to $<n_e> = 1.29\ .\ 10^9\ cm^{-3}$. The electron density in the loop region is high enough for persistent noise storm activity at metric radio wavelengths.

The high value of brightness temperature in the euv loop-region, and the positional correlation of this region with the noise storm source region, as shown in Fig. 13, suggest that there is acceleration of the non-thermal electrons along the loop, with sizes well in excess of those of the ARs that they interconnect, to energies of a few keV to a few tens of keV, and the eventual expenditure of this pent up energy (Bogod et al. 1995), about the site of the euv-brightening, as long-lasting fundamental plasma emission of the noise storm radiation.

### 5.6. Coronal Loops / Y-SXT

A systematic association of the noise storms with the global ( full - disk ) Soft X-Ray ( SXR ) flux enhancememt is known to exist, as observed from the SXR and 20 cm radio enhancements prior to the onset of noise storms (Lang and Willson 1987; Lantos 1981; Willson Lang and Liggett 1990; Raulin and Klein 1994). Metric non-thermal radio emission is known to be closely associated with flaring X-ray Bright points ( XBP ) (Kundu et al. 1994), which are persistent and of low-amplitude. XBPs are compact SXR regions associated with bipolar magnetic fields; they appear at sites of emerging magnetic flux, on the solar surface, and are hence at the foot-points of bipolar magnetic flux-tubes. Since there is unequivocal evidence for type-III-like radio bursts caused by the acceleration of non-thermal electrons along open field lines, the underlying corona may turn-up type-I continuum signature at the relatively higher observing frequencies.

The Soft X-Ray Telescope ( SXT ) (Ogawara et al. 1991; Tsuneta et al. 1991), on-board the Yohkoh Satellite, images the full disk Soft X-Ray Sun at an angular resolution of 4."92, in the 0.25 - 4 KeV energy band. The SXT is equipped with a variety of filters, and a knowledge of any pair of filter ratios is adequate in translating the imaging data to plasma emission measure and temperature parameters. The SXT full-disk images are decompressed, background dark-current, and stray light subtracted, before the plasma parameters are determined.

Spatial correlation has been employed as the criterion for associating of the SXR regions on the loop, with the source region of metric noise storms. In Fig. 14 & Fig. 15, the full-disk Yohkoh SXR image obtained with the AlMg filter ($\approx$ 0.25 − 4 $KeV$ ) at 10:22:51 UT on 2000 September 28, with an exposure duration of 5.36 s, is overlaid with the Radio Imaging contours from the GRH and the NRH. The HPBW of the radio-beams are depicted along the right side in each case. Enhanced emission in SXRs is depicted in dark shade. The SXT image on the 28th was chosen for the correlation study, on account of the non-availability of SXT data on the 26th and the 27th. Thus, the NRH total-intensity ( I ) "snapshot" image on the 28th at 11:18:11.26 UT has been overlaid on the SXT image made about one hour earlier. After correcting for differential rotation and the dissimilar image dimensions, the NRH radio-brightness contours were superposed on the Y-SXT data. The flux associated with the noise storm was 432.84 sfu ( IMP5 ( > 300 sfu ) according to (SGD-092K)).The source-size along the ($E − W \times N − S$) is (4.′75 × 5.′75). Considering the fact that the NRH has a wider beam-width along the (N-S), the peak brightness-temperature for the noise-storm source is estimated to be $7.36(\pm1.4) \times 10^8$ K.

The GRH "snapshot" of the Sun was done in an integration time of 256 ms, on the 27th, at 06:30:00 UT. The differential rotation in the SXT image was equalized, and the image dimensions were appropriately resampled by linear interpolation, in order to perform the layover with the GRH image. The calibrated solar image has a peak flux of 89 sfu, and the adjacent equispaced contour levels are apart by 9 sfu, and depict the radio brightness distribution at 109 MHz. The peak Radio brightness Temperature is 6.4 $\times$ $10^7$ K, and the adjacent contours are in multiples of $\approx$ 1.6 $\times$ $10^6$ K.

There is remarkable positional correlation of the radio source, identified as the noise s-torm continuum, with the region of enhanced SXR emission in the emission measure and temperature maps. The flaring XBPs accelerate non-thermal electrons along the reconnection loops, which are observed as the enhanced trans-equatorial features on the N-W limb of the SXT image ( their presence is hard to detect, on account of the restricted dynamic range of the Y-SXT ), while also providing them the thermal energy in the associated SXR region. The values of Total EM and temperature in the SXR region associated with the noise storm contuinuum are $7.41 \times 10^{44} \ cm^{-3}K^{-1}$ and $1.64 \times 10^{6}$ K respectively, and hence the Differential Emission Measure ( DEM ) is $6.74 \times 10^{27} \ cm^{-5}$.

The Continuum source region occurs ( as observed from the projected SXR images of Y-ohkoh ) in a region of enhanced Soft X-Rays, situated along the "legs" of the intense, and extended coronal loop (Stewart Brueckner and Dere 1986) - at an altitude of 1.22 $R_{\odot}$ in the GRH data, and 1.1 $R_{\odot}$ in the NRH data, as seen in projection, from the bipolar-footpoints; the relative positioning of the centroids of the noise-storm source with observing frequency indicates that the emission is of plasma origin. They are nearer to the apex of the connecting loop, and remain stably located within the confines of the loop, moving at the rate of a few km/s along its length (Krucker et al. 1995). The density of the SXR emitting electrons in the loop region may be determined with a knowledge of the EM. This in turn leads to the plasma frequency and plasma level in the corona wherefrom the noise storm radiation emanated. The emission measure per unit area of the loop normal to the magnetic field lines is given by Eq. 9. The observed SXR loop diameter is $d \approx (35 \pm 10)^{"}$ or $1.15 \pm 0.15 \times 10^{9} \ cm$. Assuming a loop filling factor of $q \approx 1$, and applying the DEM value estimated from the SXT map, Eq. 9 determines the mean electron density in the SXR region to be $< n_e > = (1.25 \pm 0.1) \cdot 10^{9} \ cm^{-3}$. The value of brightness temperature, as also those for $T_b$ obtained from the radio-maps of the GRH and NRH, strongly suggest the presence of a non-thermal population of electrons. The definitive frequency-characteristics of the radio-flux and $T_b$ conform the generally accepted notion that, noise storm emission occurs at the fundamental plasma emission-frequency.

## 6. Conclusions

From the observations of solar noise storm continuum events of 2000 September, 26 & 27, in the band 50 - 80 MHz, we estimated the spectral-index of the Type I continuum source to be $\approx 3.65$. The spatio-temporal and spectral correlation of the source region with various other photospheric and coronal features, as observed from the imaging data in extreme uv, soft X Ray, white-light, and $H_{\alpha}$, offers a multiwavelength perspective to the study. The close association is also borne out of the fact that, the noise storm event during this period was unique in its occurance above the NorthWest quadrant of the Sun, and hence stood a fair chance of being associated with the trans-equatorial ARs.

The height, above the photosphere, of the noise storm regions, in the case of fundamental plasma radiation, meant that the trans-equatorial loop be large; this in turn demands a corresponding increase in temperature, density, and energy content so that, they appear revealed in the midst of the multitude of loops constituting the complex magnetic topology in the ARs. Conversely, a look at the EUV and SXR data reveal the short, bright loop-structures alone (Mandrini et al. 1996; Manoharan et al. 1996), while the longer reconnection loops, which are the preferred site for the occurance of noise storms, appear either as dim apparitions or are not apparent at all. Skylab EUV and SXR images in the past have shown the existance of bright loops attaining heights of $\approx 0.1\ R_\odot$ in the corona (Švestka 1977; Dere 1982), with fainter loops capable of reaching altitudes, above the photosphere, that are nearer to the plasma-level for noise storm emission in the outer corona. In the data of SXT, EIT, and Skylab, it had been the less than adequate photon statistics ( due to a choice of smaller exposure times ) that caused the longer loops to appear much fainter than their shorter counterparts. Hence, the noise storm emission at the lower plasma frequencies, that occur at greater heights in the corona, have their associated EUV and SXR loop-regions invisible; the ones that appear are the thermally excited apex regions of shorter loops. shorter loops. Yet, the radio emission is confined to the long loops, since the density in the shorter reconnection loops are far greater for the frequencies associated with the metric noise storms that have been considered here.

The values on SXR, magnetic, radio, and optical flux, the average density of the loop region associated with noise storm activity, as well as the sunspot morphology, are in tune with the high degree of correlation existing between evolving magnetic topology in the photosphere and lower corona, the persistent generation of suprathermal electrons in the mid-corona, and their conversion to radio noise storm radiation. The metric noise storm succeeds the emergence of new magnetic flux at the site of the ARs ( slightly more than a day ), and even closer in time ( same day ) to the reconnection of magnetic fields as seen in the EUV and SXR images, while the duration of the noise storm is governed by the rate of diffusion of new flux which takes place all through the lifetime of the noise storm. The fact that this duration encompasses the life-span of quiet a few coronal-loops suggests that, the noise storm source region is brought into being by a system ( or an arcade ) of loops. The estimation of plasma parameters like brightness temperature, suprathermal electron beam-density, Emission-Measure, and quantities like radio-flux, radio brightness-temperature, fluctuations in magnetic flux-density in the active regions, unequivocally establish a fundamental plasma emission mechanism for the continuum noise storm radiation, brought about by the energet-ics that link the magnetic flux to the non-thermal electrons in the corona.

We thank Dr. R. Ramesh for useful discussions, and E. Ebenezer for the preliminary data reduction performed, in connection with this work. We also thank the referee for insightful comments which improved the content and presentation of this paper. We also wish to express our gratitude to the staff of the Gauribidanur Radio Observatory, as well as the Nançay RadioHeliograph team at the Unite Scientifique de Nançay ( Station de RadioAstronomie de Nançay, 18330 Nançay, France ) of the Observatoire de Paris, for use of their 2D radio maps. The Solar Geophysical Data Reports are published by the National Geophysical Data Center, NOAA, 325 Broadway, Boulder, Colorado, 80305-3328 USA. The $H_\alpha$ images were obtained from the FTP Data Archive and WWW pages of the Big Bear Solar Observatory/New Jersey Insitute of Technology, 40386 North Shore Lane, Big Bear City, CA 92314 USA. The SoHO spacecraft, with the LASCO, MDI, EIT instruments on board, is an international joint venture involving ESA and NASA. The solar X-ray image is from the Yohkoh mission of ISAS, Japan. The X-ray telescope was prepared by the Lockheed Palo Alto Research Laboratory, the National Astronomical Observatory of Japan, and the University of Tokyo with the support of NASA and ISAS. SolarSoftWare (SSW) system is built from Yohkoh, SoHO, SDAC, and Astronomy libraries, draws upon contributions from many members of those projects, and is made available at the Lockheed Martin Solar and Astrophysics Laboratory ( LMSAL ) website. CHIANTI is a collaborative project involving the Naval Research Laboratory (Washington DC, USA), the Arcetri Observatory (Firenze, Italy), and the Cambridge University, UK.